\documentclass[twocolumn,showpacs,preprintnumbers,amsmath,amssymb,showkeys]{revtex4}

\usepackage{graphicx}
\usepackage{dcolumn}
\usepackage{bm}

\begin{document}

\preprint{APS/123-QED}

\title{Spin accumulation caused by confining potential \\in a two dimensional electron system}

\author{Kazuhisa Suzuki}
\email{kazuhisa@kh.phys.waseda.ac.jp}
\author{Susumu Kurihara}
\affiliation{Department of physics, Waseda University, Okubo, Shinjuku, Tokyo 169-8555, Japan.}
\date{\today}

\begin{abstract}
We investigate spin accumulation caused by spin-orbit interactions (SOIs) in a two dimensional electron system confined in one direction.
 We calculate spin density caused by three kinds of SOIs; arising from edge potential, Rashba and Dresselhaus mechanisms. All SOIs are shown to generate out-of-plane spin accumulations at the lateral edges of the system. We also find that the Rashba and the Dresselhaus mechanisms are competitive. Especially, when the strength of both interactions are equal, their effects cancel out each other. In that case, only the edge potential mechanism becomes relevant, and analytical expression is obtained for the spin density. The edge potential mechanism is shown to induce spin accumulation similar to and consistent with the experiment [Sih {\it et al}., Nature Phys {\bf 1}, 31 (2005)]. We discuss the mechanism of the spin accumulation for each SOIs in some detail.
\end{abstract}

\maketitle

\section{\label{sec:level1}Introduction} 

Spin-orbit interaction (SOI) in semiconductors attracts much attention, since it makes possible to control electron spins without magnetic field, a goal of spintronics \cite{prinz, wolf}. SOI is very small in a vacuum, since its strength is inversely proportional to the Dirac gap creating electron-positron pairs $\sim$ $10^{6}$ eV. In contrast to a vacuum, SOI is largely enchanced for electrons in semiconductors, since the corresponding energy of creating electron-hole paris is given by the band gap $\Delta\sim1$ eV \cite{rashba}. This is why SOIs play important roles in spintronics.

SOIs have been shown to generate spin currents perpendicular to bias current, a phenomenon called spin Hall effect (SHE), which is expected to be a useful tool for creating spin polarized current. SHE is explained as a result of spin dependent sccaterings through impurity potential by D'yakonov and Perel' \cite{dp}, and independently by Hirsch \cite{hirsch} and Zhang \cite{zhang}. 
Meanwhile, other scenarios are proposed for $p-$type bulk semiconductors described by Luttinger Hamiltonian \cite{murakami}, and for $n-$type two dimensional semiconductors with the Rashba type SOI \cite{sinova}. In addition to the Rashba SOI \cite{rashbahami}, Dresselhaus SOI \cite{drehami} is also shown to generate spin current in exactly opposite direction to the one which the Rashba SOI gives \cite{sinitsyn}.

These transverse spin currents have been expected to generate opposite out-of-plane spin accumulations at the lateral edges of sample, analogously to Hall effect; longitudinal current induces charges with opposite signs at the transverse edges. This theoretical prediction was confirmed in the recent experiments \cite{kato, wunderlich, sih, stern}. Several theoretical studies suggest that the spin accumulation are caused by three SOIs \cite{governale, nikolic1, qwang, jwang, malshukov, usaj, yao, xing}. The Rashba SOI \cite{governale, nikolic1, qwang, jwang, usaj, yao} and the Dresselhaus SOI \cite{malshukov} are shown to cause opposite spin accumulations at lateral edges. Recently, Xing {\it et al}. \cite{xing} have shown that the SOI coming from the edge potential can induce the spin accumulation as well. Here, a question arises, as to how these three SOIs determine spin density. As far as authors are aware, no calculation including all three types of SOIs has been published.
   
In this paper we study the effects of three SOIs on spin accumulation in a two dimensional electron system (2DES). Spin density is calculated analytically for the SOI from edge potential, and numerically for the Rashba and the Dresselhaus SOIs. Finally we investigate the effects of all SOIs. All SOIs are shown to generate opposite out-of-plane spin accumulations at the lateral edges. We also find that the Rashba and the Dresselhaus SOIs are competitive, as is the case for SHE. The SOI from the edge potential is shown to generate similar acuumulation to the experiment \cite{sih}. For each SOIs we discuss precisely the mechanism of spin accumulation.

\begin{figure}[t]
\includegraphics[width=60mm, scale=1]{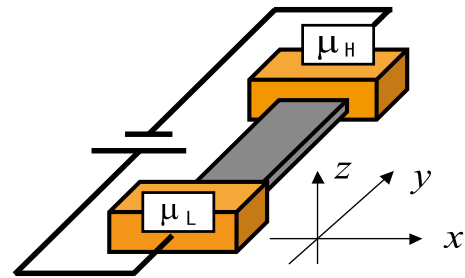}
\caption{\label{fig:epsart} Schematic figure of the system. A two dimensional system is sandwiched between two contacts. $\mu_{\rm H}$ and $\mu_{\rm L}$ are chemical potentials and we assume $\mu_{\rm H}$ \textgreater $\mu_{\rm L}$.}
\label{fig:system}
\end{figure}
\section{Model Hamiltonian}
\begin{figure}[t]
\includegraphics[width=85mm, scale=1]{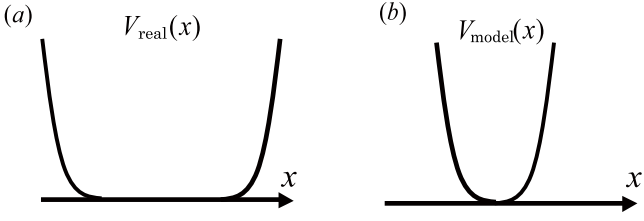}
\caption{\label{fig:epsart} (a) The edge potential in real systems: $V_{\rm real}(x)$ is zero in the center and gradually increases toward the edges. (b) The edge potential in our study: considering only the edge effect}
\label{fig:edgepotential}
\end{figure}

 We consider a ballistic 2DES with two reflectionless contacts, illustrated in Fig.\ref{fig:system}. It means that electrons injected from one contact move freely without scatterings in the 2DES, and enter the other contact without reflections. In our setup, electrons are confined in $x$ $(z)$ direction by edge potential $V_{\rm edge}(x)$ $(V(z))$. Although the edge potential profile in a real system is like $V_{\rm real}(x)$ in Fig.\ref{fig:edgepotential}(a), considering the edge potential like $V_{\rm model}(x)$ in Fig.\ref{fig:edgepotential}(b) is enough to understand the spin accumulation observed in the experiments. It is because the spin accumulation occurs only in the vicinity of the edges, and as far as the edge spin polarization is concerned, there is no essential difference between these two potentials.
 We assume $V_{\rm model}(x)$ to be harmonic potential to make calculation easier. In this model, Hamiltonian for electrons compose of free electron part $H_{0}$ and the SOIs part $H_{\rm SOI}$. Hamiltonian for free electrons is given by 
\begin{eqnarray}
&&H_{0} = \frac{1}{2m^*}(p^{2}_{x}+p^{2}_{y})+V_{\rm model}(x)~,\\
&&V_{\rm model}=\frac{1}{2}m^*\omega^2x^2~,
\end{eqnarray} 
where $\overrightarrow{p}$ is momentum, $m^*$ is effective mass of electrons (in our work, we choose GaAs as the host of the electronic syetem, i.e. $m^*=0.067m_{e}$), $\omega$ is a harmonic oscillator frequency. We assume that the confining potential $V(z)$ is so strong that all electrons are in the lowest energy state in $z$ direction.
Hamiltonian for the SOIs consists of three parts,
\begin{eqnarray}
&&H_{\rm SOI}=H_{\rm Edge}+H_{\rm R}+H_{\rm D}~, \\
&&H_{\rm edge}=\frac{(g^*-1)\hbar}{2m^*\Delta}\sigma_{z}p_{y}\frac{dV_{\rm model}(x)}{dx}~, \\
&&H_{\rm R}=\frac{\alpha}{\hbar}(p_{x}\sigma_{y}-p_{y}\sigma_{x})~, \\
&&H_{\rm D}=\frac{\beta}{\hbar}(p_{x}\sigma_{x}-p_{y}\sigma_{y})~,
\end{eqnarray}
where $H_{\rm edge}$, $H_{\rm R}$ and $H_{\rm D}$ are respectively the SOI from the edge potential, the Rashba SOI, and the Dresselhaus SOI. 
$\sigma_{i}~(i=x,y,z)$ is Pauli matrix, $g^*$ is Lande factor ($g^*=-0.44$ in GaAs), $\hbar$ is Planck constant divided by $2\pi$, $\alpha$ and $\beta$ characterize the strength of the Rashba and the Dresselhaus SOIs, $\Delta$ is $1.42$ eV in GaAs. Hereafter we take $z$ axis as the spin quantization axis.  

\section{Calculations and Results}
In this section, we calculate spin density in $z$ direction $\langle\sigma_{z}\rangle$ which was detected in the experiments \cite{kato, wunderlich, sih, stern}. To investigate the effects of each SOIs precisely, we divide this section into three subsections. Firstly we study the effect of only the SOI from the edge potential, and secondly the Rashba and the Dresselhaus SOIs, and finally, all of them. All calculations are done assuming zero temperature.
\subsection{SOI from the edge potential }

In the case of the SOI from the edge potential, Hamiltonian is given,
\begin{equation}
H=\frac{1}{2m^*}(p^{2}_{x}+p^{2}_{y})+\frac{1}{2}m^{*}\omega^2x^2+\frac{(g^*-1)\hbar\omega^2}{\Delta}\sigma_{z}p_{y}x~.
\end{equation}
\begin{figure}[t]
\includegraphics[width=60mm, scale=1]{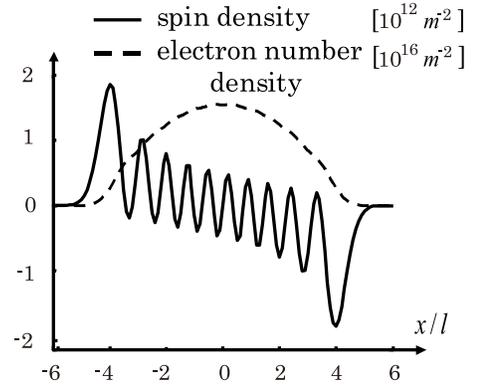}
\caption{\label{fig:epsart} Spin density and electron density are plotted as a function of x for Fermi energy $E_{\rm F}=50$ meV, $\hbar\omega=5$ meV, $\mu_{H}-\mu_{L}=1$ meV. Up-spin (down-spin) electrons accumulate in $-$ $(+)$ $x$ direction. $l=\sqrt{\hbar/m^{*}\omega}$ is the characteristic length of the harmonic oscillator.}
\label{fig:cond2}
\end{figure}
Here, we can replace $\sigma_{z}$ and $p_{y}$ respectively, by $s$ ($+1$ for up-spin, $-1$ for down-spin) and $\hbar k_{y}$, since these two operators commute with $H$. Before diagonalization, we rewrite Hamiltonian to see how the edge potential works on electrons through SOI,
\begin{eqnarray}
H=\frac{p^{2}_{x}}{2m^*}+\frac{1}{2}m^{*}\omega^2(x+\frac{(g^*-1)\hbar^2}{2m^*\Delta}sk_{y})^2 \nonumber \\
+\frac{\hbar^2}{2m^*}(1+\frac{1}{2}(g^*-1)^2(\frac{\hbar\omega}{\Delta})^2)k_{y}^2~,
\end{eqnarray}
The second term represents harmonic potential depending on spin and wavenumber. Up- and down-spin electrons with the same wavenumber $k_{y}$ feel different potential effectively. For example, for up-spin (down-spin) electron with $-k_{y}$, its potential minimum shifts to $-$ $(+)$ $x$ direction. This Hamiltonian can be easily diagonalized and eigenstates and energies are given by

\begin{eqnarray}
&&\Psi_{n,k_{y},s}(x,y)=\frac{e^{i k_{y} y}}{\sqrt{L_{y}}}\Phi_{n}(x+\frac{(g^*-1)\hbar^2}{2m^*\Delta}sk_{y})\chi_{\rm s}~, \\
&&E_{n,k_{y},s}=\hbar\omega(n+\frac{1}{2}) \nonumber \\
&& \qquad \qquad+\frac{\hbar^2}{2m^*}(1+\frac{1}{2}(g^*-1)^2(\frac{\hbar\omega}{\Delta})^2)k_{y}^2~,
\end{eqnarray}
where $\Phi_{n}(x)$ and $\chi_{s}$ are wavefunctions of the harmonic oscillator for band index $n$ and of spin part. As we can see easily, the center of density profile shifts depending on the electron spin in $z$ direction and the wavenumber in y direction. Up- and down-spin states are energetically degenerate. Using these wavefunctions, we calculate spin density $\langle\sigma_{z}\rangle$ summing up the expectation value for occupied state $(n,k_{y},s)$,
\begin{equation}
\langle\sigma_{z}\rangle=\sum_{n,k_{y},s}\Psi_{n,k_{y},s}^{\dagger}(x,y)\sigma_{z}\Psi_{n,k_{y},s}(x,y)~.
\end{equation}
An assumption of reflectionless contacts greatly simplifies the treatment of applied electric field: in Fig.\ref{fig:system}, electrons with positive (negative) momentum in y direction occupy the states with energy lower than $\mu_{\rm L}$ $(\mu_{\rm H})$, since electrons injected from one contact go through the other side contact without reflections.
In our assumption $\mu_{\rm H} > \mu_{\rm L}$, electrons with energy $\mu_{\rm L} < E < \mu_{\rm H}$ have negative wavenumber, and contribute to spin accumulation at zero temperature.
By assuming $\mu_{\rm H}-\mu_{\rm L} \ll \mu_{\rm H},\mu_{\rm L}$, we obtain analytical expression of spin density in the linear response regime,
\begin{eqnarray}
&&\langle\sigma_{z}\rangle=\sum_{n}\frac{mL_{y}}{2\pi\hbar^2(-k_{y}^{\rm F}(n))}(|\Psi_{n,k_{y}^{\rm F}(n),s=1}(x)|^2 \nonumber \\
&&\qquad\qquad\quad -|\Psi_{n,k_{y}^{\rm F}(n),s=-1}(x)|^2)(\mu_{H}-\mu_{L})~, 
\end{eqnarray}
where $k_{y}^{\rm F}(n)$ is Fermi wavenumber of band index {\it n} given by 
\begin{equation}
k_{y}^{\rm F}(n)=-\sqrt{\frac{E_{F}-\hbar\omega(n+\frac{1}{2})}{\frac{\hbar^2}{2m^*}(1+\frac{1}{2}(g^*-1)^2(\frac{\hbar\omega}{\Delta})^2)}}~.
\end{equation}

\begin{figure}[t]
\includegraphics[width=60mm, scale=1]{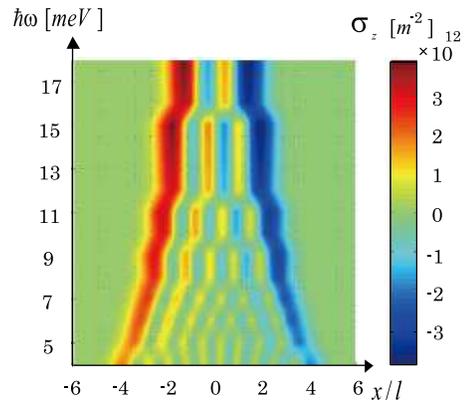}
\caption{\label{fig:epsart} Spin density is plotted as a function of $x$ and the edge potential strength $\hbar\omega$ for $E_{\rm F}=50$ meV, $\mu_{H}-\mu_{L}=1$ meV. The width of the spin density and the numbers of oscillations depend on the trap strength.}
\label{fig:edge1}
\end{figure}

In this expression, $|\Psi_{n,k_{y}^{\rm F}(n),s=+(-)1}(x)|^2$ represents that up-spin (down-spin) electron accumulates in $-$ $(+)$ $x$ direction, so contributions from two states $(n,k_{y}^{\rm F}(n),s=\pm1)$ result in opposite spin accumulations. Of course, the summation for band index {\it n} gives the similar result. The spin density and the electron number density are plotted in Fig.\ref{fig:cond2}. From this figure, it is seen that the SOI from the edge potential causes opposite spin accumulations at the lateral edges. The spin density profile depends on the Lande factor $g^{*} \gtrless 1$. In the case $g^{*} > 1$, spin density profile is opposite to the result in Fig.\ref{fig:cond2}. We also plot the spin density, as a function of the strength of the harmonic potential with keeping Fermi energy constant in Fig.\ref{fig:edge1}. The width of the spin density and the number of its oscillations are characterized by the wavefunctions of electrons in the highest band. In the strong trapping limit, only electrons in the lowest subband generate the spin accumulation with the narrow width and a small number of oscillations. As the potential become weaker, the more subbands are involved and the spin density oscillations spreads over the wide region. The step like change occurs when the number of subbands crossing the Fermi level changes. Spin polarization, $P=\langle \sigma_{z} \rangle/\langle n \rangle$, at the peak of the spin accumulation is about $10^{-3}\sim10^{-4}$ in Fig.\ref{fig:cond2} and Fig.\ref{fig:edge1}. 
\begin{figure}[t]
\includegraphics[width=60mm, scale=1]{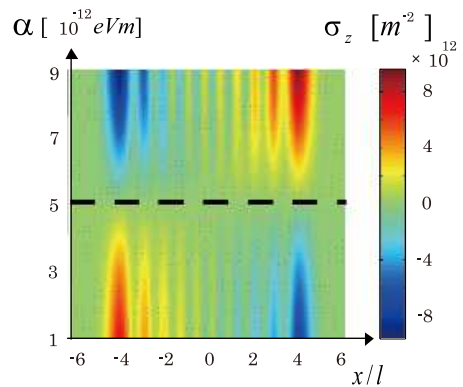}
\caption{\label{fig:epsart} Spin density is plotted as a function of $x$ and $\alpha$ for $E_{\rm F}=50$ meV, $\hbar\omega=5$ meV, $\mu_{H}-\mu_{L}=1$ meV, and  $\beta =5\times 10^{-12}$ eVm. The Rashba and the Dresselhaus SOIs are competitive. Dash line indicates $\alpha=\beta$.}
\label{fig:rashdre}
\end{figure}
\subsection{Rashba SOI and Dresselhaus SOI}
We treat the Rashba and the Dresselhaus SOIs on an equal footing since they are theoretically shown to generate spin currents in opposite directions with the same magnitude, and expected to give similar effects on spin density. In our model, Hamiltonian is given by,
\begin{eqnarray}
H&=&\frac{1}{2m^*}(p^{2}_{x}+p^{2}_{y})+\frac{1}{2}m^{*}\omega^2x^2 \nonumber \\
&+&\frac{\alpha}{\hbar}(p_{x}\sigma_{y}-p_{y}\sigma_{x})+\frac{\beta}{\hbar}(p_{x}\sigma_{x}-p_{y}\sigma_{y})~,
\end{eqnarray}
Similar to the previous section, we can put $p_{y}=\hbar k_{y}$. Wavefunctions is expanded in eigenfunctions of the harmonic oscillator,
\begin{equation}
\tilde{\Psi}_{N,k_{y},\pm}(x,y)=\sum_{n,k_{y},s}a_{n,s}^{(N,\pm)}\Psi_{n,k_{y},s}~,
\end{equation}
\noindent
where {\it N} is a new band index, $\pm$ indicate energy bands split caused by the SOIs. Coefficients $a_{n,s}^{(N,\pm)}$ are determined numerically. We expect non-zero spin density in applied electric field as in the previous section. A calculation for spin density $\langle\sigma_{z}\rangle$ is the same as Eq$(11)$. To see the effects of two SOIs, the spin density is plotted in Fig.\ref{fig:rashdre}, as a function of the Rashba strength with keeping the Dresselhaus strength constant, since the Rashba strength can be tuned by gate voltage \cite{nitta}. From Fig.\ref{fig:rashdre}, two SOIs are shown to generate competitive result, that for $\alpha>\beta$ ($\alpha<\beta$), the spin density at the edges is up-spin-like (down-spin-like) in $+x$ direction, and down-spin-like (up-spin-like) in $-x$ direction. In the case $\alpha=\beta$, the effects of two SOIs cancel out each other and the spin density is exactly zero in the entire region. Spin polarization at the peak is about $10^{-3}$. This accumulation is obviously caused by two SOIs, but how the SOIs work physically is unclear. Such a difficulty comes from the fact that electron spins are not conserved. To see how the SOIs change electron spin states, we derive dynamical force acting on electrons in $x$ direction, given by $F_{x}=m^{*}d^2x/d^2t=m^{*}(i\hbar)^{-2}[[x,H],H]$ \cite{nikolic},
\begin{equation}
F_{x}=-\frac{dV_{\rm Edge}(x)}{dx}+\frac{2m^*}{\hbar^2}(\alpha^2-\beta^2)k_{y}\sigma_{z}~.
\end{equation}
The second term shows that the Rashba and the Dresselhaus SOIs cancel out in the case $|\alpha|=|\beta|$. In quantum mechanics, the force is hard to deal with, so we calculate back an effective potential from $F_{x}$, just as in classical mechanics, giving a spin dependent effective potential, 
\begin{eqnarray}
V_{\rm eff}(x)&=&-\int F_{x} dx  \nonumber \\
 &=&V_{\rm Edge}(x)-\frac{2m^*}{\hbar^2}(\alpha^2-\beta^2)k_{y}\sigma_{z}x~.
\end{eqnarray}
From this expression, it is clear that the second term determins the profile of spin density.
\begin{figure}[t]
\includegraphics[width=70mm, scale=1]{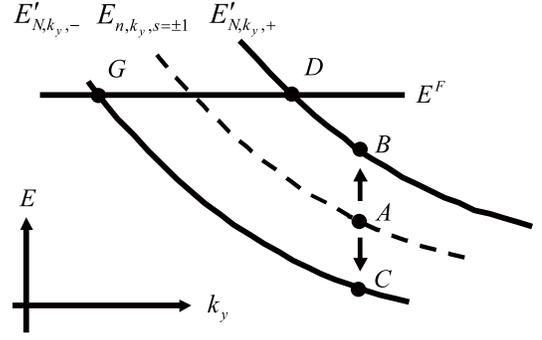}
\caption{\label{fig:epsart} Schematic figure of energy separation. The degenerate state A with energy $E_{n,k_{y},\pm1}$, eigenenrgy of $H_{0}$, is separate into the stete B with higher energy $E_{n,k_{y},+}^{'}$ and the state C with lower energy $E_{n,k_{y},-}^{'}$. D and G are the states with Fermi energy.}
\label{fig:separation}
\end{figure}
\begin{figure}[t]
\includegraphics[width=70mm, scale=1]{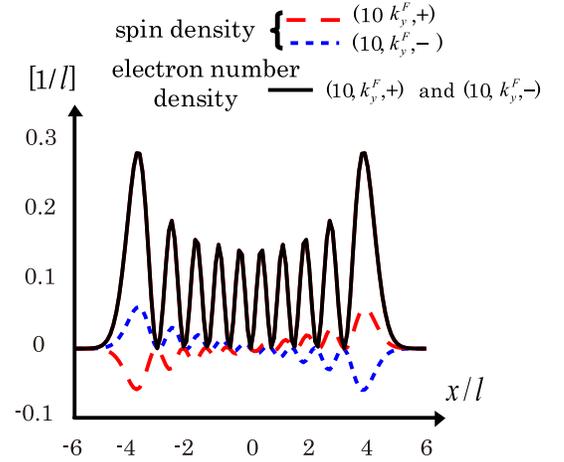}
\caption{\label{fig:epsart} Electron densities and spin densities are plotted as a function of x for state $(10,k_{y}^{\rm F},+)$ and $(10,k_{y}^{\rm F},-)$ in highest two bands crossing Fermi energy. Parameters are $E_{\rm F}=50$ meV, $\hbar\omega=5$ meV, $\alpha =7\times 10^{-12}$ eVm, $\beta =5\times 10^{-12}$ eVm. Solid line represents that the profiles of electron number densities are approximately same. Dashed line and dotted line show the doformed spin densities for two state caused by SOIs.} 
\label{fig:secondterm}
\end{figure}

We now consider energy bands to discuss the effect of the second term, as illustrated in Fig.\ref{fig:separation}. A state A with energy $E_{n,k_{y},s=\pm1}$, an eigenvalue of Hamiltonian $H_{0}$, is degenerate for up- and down-spin. Spin density, contributed from the states $(n,k_{y},s=1)$ and $(n,k_{y},s=-1)$, is exactly zero, since spin densities for up- and down-spin have the same profile with opposite signs, and cancel out each other. The SOIs lift the degeneracy of two states into a state B with higher enegy $E_{N,k_{y},+}^{'}$ and a state C with lower energy $E_{N,k_{y},-}^{'}$. In the case $k_{y}\textless0$ and $|\alpha|\textgreater|\beta|$, for the state B (C) , the spin density, following the second term, behaves up-spin-like in $x>0$ $(x<0)$ and down-spin-like in $x<0$ $(x>0)$, while the electron density is approximately unchanged. In Fig.\ref{fig:secondterm}, we plot the spin densities and the electron number densities for two states, say states D and G. Here the states D and G are taken to be the two highest band states at the Fermi level. In our parameter, they correspond to the states labeled by $(10,k_{y}^{\rm F},+)$ and $(10,k_{y}^{\rm F},-)$. Profiles of electron densities for two states are nealy the same (solid line), but profiles of spin densities are different as explained above (dashed and dotted line). In our system, the density of states is inversely proportional to the square root of energy measured from the bottom of subband. With the same band index {\it N} and energy {\it E}, electrons in $+$ band contribute to spin density more than electrons in $-$ band do, then the result for $|\alpha|\textgreater|\beta|$ in Fig.\ref{fig:rashdre} is similar to the spin density profile for electrons in $+$ bands. In this way, we can explain the induced the spin accumulation with the discussion from the deformation of the spin density caused by the SOIs and the density of states.

\subsection{All SOIs}
\begin{figure}[t]
\includegraphics[width=60mm, scale=1]{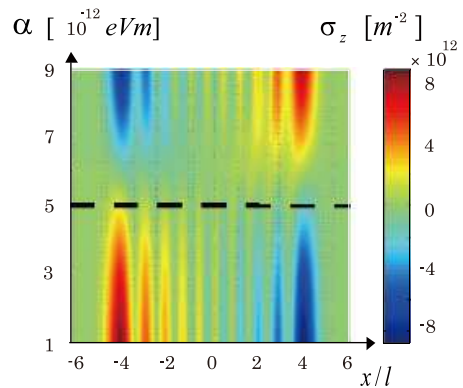}
\caption{\label{fig:epsart} Spin density caused by three SOIs is plotted as a function of $x$ and $\alpha$ for $E_{\rm F}=50$ meV, $\hbar\omega=5$ meV, $\mu_{H}-\mu_{L}=1$ meV, and $\beta =5\times10^{-12}$ eVm. Spin density is similar to the previous result in Fig.\ref{fig:rashdre}, but for $|\alpha|=|\beta|$ the spin density is not zero because of the SOI from the edge potential.}
\label{fig:edgerashdre}
\end{figure}

In this subsection, we consider all SOIs. Hamiltonian is exactly diagonalized numerically by expanding wavefunctions in eigenstates of the harmonic oscillator. Calculated spin density $\langle\sigma_{z}\rangle$ is plotted in Fig.\ref{fig:edgerashdre}, representing the effects of three SOIs. This figure looks similar to the previous result (Fig.\ref{fig:rashdre}), but, even in the case $|\alpha|=|\beta|$, the spin density is not zero. It is because the SOI from the edge potential makes spin density at the edges up-spin-like (down-spin-like) in $-$ $(+)$ $x$  direction. In that case, the spin density profile is the exactly same result of only the SOI from the edge potential, as long as $|\alpha|=|\beta|$. We believe that a result obtained in the realistic potential like Fig.\ref{fig:edgepotential}(a) is essentially same with our results, since only the physical mecanism at the edges is important.

We obtain similar spin accumulation to the experiment ($\alpha=1.8\times10^{-12}$ eVm, $\beta=0$ eVm), performed in the two dimensional electron system \cite{sih}. In that experiment, the observed spin density is up-spin-like (down-spin-like) at the edges in $-$ $(+)$ $x$ direction. This shows the importance of the SOI from the edge potential, since the theory taking account of the Rashba SOI alone can not explain the experimental result \cite{sih}. 

One of the most characteristic results is that the signs of spin accumulation from the edge potential is opposite depending on the Lande factor $g^{*} \gtrless 1$. It is because the sign of $H_{\rm edge}$ $(4)$ depends on the Lande factor. Actually, in the experiment performed in ZnSe ($g^{*}=1.1$) \cite{stern}, the sign of the spin accumulation is opposite to the experiment peformed in GaAs ($g^{*}=-0.44$) \cite{kato, sih}. These experimental results are consistent with our analysis, and demonstrate the important role of the SOI from the edge potential. 
\section{Summary}
We have investigated the spin accumulation caused by three SOIs in the 2DES. The spin density $\langle\sigma_{z}\rangle$ was calculated including all SOIs. All three SOIs were shown to induce opposite spin accumulations at the lateral edges, which was observed in the experiments . We also showed that the Rashba and the Dresselhaus SOIs produce competitive results. Especially when the strength of both SOIs are equal, their effects cancel each other. In that case, the spin accumulation caused by the SOI from the edge potential remains, and the analytical expression was obtained for the spin density. In addition, we discussed precisely the physical mechanism of forming opposite spin accumulations for each SOIs, considering effective potentials acting on electrons. Finally, we discussed our results, compared with the experiments, and concluded that the SOI from the edge potential induces the observed spin accumulation.
From our result, we predict that the spin accumulation for $g^{*}<1$ ($g^{*}>1$) can be diminished (enhanced) as the Rashba strength increase, which can be observed by tuning the gate voltage. This electrical control of electron spins would become a useful tool in spintronics.

\begin{acknowledgments}
We thank K. Kamide and N. Yokoshi for valuable comments and discussions.
 This work is partly supported by The 21th Century COE Program at Waseda University from the Ministry of Education, Sports, Culture, Science and Technology of Japan.
\end{acknowledgments}

\newpage 

%

\end{document}